\documentclass[aps,prb,showpacs,twocolumn,floats]{revtex4}
\usepackage{amssymb}

\usepackage{amsmath}
\usepackage{graphicx,psfig}
\usepackage{dcolumn}
\usepackage{bm}

\input{tcilatex}

\begin{document}

\title{Deflagration with quantum and dipolar effects in a model of a molecular
magnet}
\author{D. A. Garanin, Reem Jaafar}
\affiliation{ \mbox{Department of Physics and Astronomy, Lehman College, City
University of New York,} \\ \mbox{250 Bedford Park Boulevard
West, Bronx, New York 10468-1589, U.S.A.} }
\date{10 February 2010}

\begin{abstract}
Combination of the thermal effet in magnetic deflagration with resonance
spin tunneling controlled by the dipole-dipole interaction in molecular
magnets leads to the increase of the deflagration speed in the \emph{dipolar
window} near tunneling resonances.
\end{abstract}
\maketitle

\begin{figure}[th]
\unitlength1cm
\begin{picture}(11,5)
\psfig{file=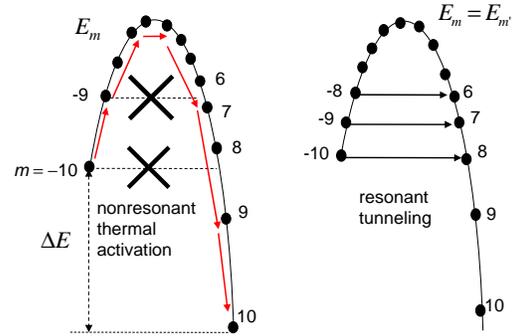,angle=-90,width=8cm}
\end{picture}
\caption{Thermal activation in the nonresonant case vs tunneling in the
resonant case. }
\label{Fig-Levels}
\end{figure}

Molecular magnets such as Mn$_{12}$ Ac, possessing an effective large spin $S
$=10, are famous as mesoscopic systems demonstrating magnetic bistability
due to the strong uniaxial anisotropy. Spectacular spin tunneling under the
anisotropy barrier in molecular magnets was first seen in the steps in
dynamical hysteresis curves. \cite{frisartejzio96prl} The steps correspond
to the values of the longitudinal magnetic field $B_{z}$ at which the energy
levels in both potential wells match. Here resonance spin tunneling leads to
a faster relaxation responsible for a step of the magnetization. To the
contrast, off resonance the main channel of relaxation is thermal activation
over the top of the barrier. The difference between the two types of
relaxation is shown in Fig. \ref{Fig-Levels}. In fact, spin tunneling
requires a transverse field or any other term in the Hamiltonian that breaks
the axial symmetry. Pure spin tunneling in the right panel in Fig. \ref
{Fig-Levels} requires that these terms be sufficiently strong, such as the
transverse field of about 3 T in Mn$_{12}.$ In the case of weaker tunneling
interactions, the intermediate situation of a thermally assisted tunneling
is realized. In this case spins tunnel after thermally mounting up to below
the top of the barrier. \cite
{garchu97prb,chugar97prl,garmarchu98prb,garchu99prb} The role of tunneling
in the case of weaker tunneling interactions can be interpreted as some
lowering of the barrier near resonances. For stronger tunneling
interactions, the barrier is removed completely at resonances.

Tunneling and relaxation in molecular magnets can be described by the
density matrix equation, \cite{garchu97prb} the most comprehensive account
of which is given in Ref. \onlinecite{gar08-DME}. The latter numerically
implements the universal spin-phonon interaction suggested in Refs.
\onlinecite
{chu04prl,chugarsch05prb}. This interaction is due to distortionless
rotation of the crystal field acting on the spins by transverse phonons and
it is completely expressed in terms of the crystal-field Hamiltonian $\hat{H}%
_{A}$ without any unknown spin-lattice coupling constants.

Experiments of 2005 by Myriam Sarachik group showed the existence of
propagating deflagration (burning) fronts in the molecular magnet Mn$_{12}$
Ac that are similar to chemical burning. \cite{suzetal05prl,heretal05prl}
Javier Tejada group observed peaks in the deflagration speed on the bias
magnetic field $B_{z}$ that were interpreted as contribution of resonance
spin tunneling.\cite{heretal05prl} A detailed, mainly classical, theory of
the magnetic deflagration including the ignition threshold and the accurate
prefactor in the Arrhenius-type expression for the speed of the burning
front was proposed in Ref. \onlinecite
{garchu07prb}.

The physics of deflagration is based on triggering relaxation from a
metastable state over potential barrier by the temperature increase as the
result of relaxation accompanied by energy release. The burning front forms
because the temperature in the regions still unburned (e.g., before the
front) rises as the result of heat conduction from the hot areas where
burning just occured. The two main ingredients of deflagration thus are the
Arrhenius dependence of the relaxation rate on temperature (making burning
in the cold areas before the front negligibly slow) and heat conduction.
Deflagration is mathematically described by the system of coupled i) rate
equation for the number of particles (magnetic molecules) in the metastable
state and ii) the heat conduction equation.

Subsequent theoretical quest for an essentially quantum form of deflagration
lead to the discovery of self-organized fronts of tunneling, a non-thermal
process triggered by the dipolar field (rather than by temperature) that can
bring the system on or off resonance. \cite{garchu08prb,garchu09prl,gar09prb}
A hallmark of these fronts is the self-consistent adjustment of the
metastable population (or magnetization) to the optimal spatial profile that
creates the dipolar field that is constant in some region of space and
brings the system on resonance. The width of the resonance region forming
the front core is about the transverse dimension $R$ of the sample that
allows an efficient tunneling and thus front propagation. On the other hand,
before and after the front core the system is off resonance and tunneling is
blocked. Fronts of tunneling can be realized in the \emph{dipolar window} of
the external field $B_{z}$
\begin{equation}
0\leq B_{z}-B_{k}\leq B_{z}^{(D)}.  \label{DipolarWindow}
\end{equation}
Here $B_{k}$ is the field corresponding to the $k$th resonance in the
absence of the dipolar field and $B_{z}^{(D)}$ is the dipolar field created
by the uniformly magnetized elongated crystal. It was shown \cite
{garchu09prl} that the adjustment mechanism is robust with respect to
resonance spread (e.g., due to defects) smaller than $B_{z}^{(D)}.$

The aim of the present paper is to unify the theories of the standard (hot)
deflagration\cite{suzetal05prl,garchu97prb} and fronts of tunneling (cold or
quantum deflagration).\cite{garchu09prl,gar09prb}

In the sequel we will use the generic model of a molecular magnet with $%
\hat{H}_{A}=-DS_{z}^{2},$ where the tunneling resonance fields are given by
\begin{equation}
B_{k}=kD/\left( g\mu _{B}\right) ,\qquad k=0,1,\ldots   \label{BkDef}
\end{equation}
Resonance tunneling occurs at $B_{\mathrm{tot},z}=B_{z}+B_{z}^{(D)}\approx
B_{k}$ between the metastable ground state $\left| -S\right\rangle $ and an
excited state at the other side of the barier $\left| m^{\prime
}\right\rangle $ with $m^{\prime }=S-k.$ At temperatures much smaller than
the barrier height (e.g. at the temperature of the deflagration front) one
can describe magnetic molecules as two-level systems occupying the states $%
\left| \pm S\right\rangle .$ Let us denote the probability for a molecule to
be in the metastable state $\left| -S\right\rangle $ as $n.$ Then the
average value of the effective pseudospin $\sigma _{z}\ $is
\begin{equation}
\sigma _{z}=1-2n,  \label{sigmazAvrOverd}
\end{equation}
so that $n=1$ corresponds to $\sigma _{z}=-1.$ The general expression for
the longitudinal component of the dipolar field on magnetic molecule $i$ is
the sum over positions of all other moleculs $j$%
\begin{equation}
B_{i,z}^{(D)}=\frac{Sg\mu _{B}}{v_{0}}D_{i,zz},\qquad D_{i,zz}\equiv
\sum_{j}\phi _{ij}\sigma _{jz}.  \label{BviaD}
\end{equation}
Here $v_{0}$ is the unit-cell volume, $D_{zz}$ is the reduced dipolar field,
and
\begin{equation}
\phi _{ij}=v_{0}\frac{3\left( \mathbf{e}_{z}\cdot \mathbf{n}_{ij}\right)
^{2}-1}{r_{ij}^{3}},\qquad \mathbf{n}_{ij}\equiv \frac{\mathbf{r}_{ij}}{%
r_{ij}}.  \label{psiijDef}
\end{equation}
Inside a uniformly magnetized ellipsoid, $\sigma _{z}=\mathrm{const},$ the
dipolar field is uniform and one has $D_{zz}=\bar{D}_{zz}\sigma _{z},$ where
\begin{equation}
\bar{D}_{zz}=\bar{D}_{zz}^{(\mathrm{sph})}+4\pi \nu \left(
1/3-n^{(z)}\right) ,  \label{Dzzbar}
\end{equation}
$\nu $ is the number of magnetic molecules per unit cell ($\nu =2$ for Mn$%
_{12}$ Ac) and $n^{(z)}=0,$ $1/3,$ and 1 for a cylinder, sphere, and disc,
respectively. The reduced dipolar field in a sphere $\bar{D}_{zz}^{(\mathrm{%
sph})}$ depends on the lattice structure. For Mn$_{12}$ Ac lattice summation
yields $\bar{D}_{zz}^{(\mathrm{sph})}=2.155$ that results in $\bar{D}_{zz}^{(%
\mathrm{cyl})}=10.53$ for a cylinder. Then Eq.\ (\ref{BviaD}) yields the
dipolar field $B_{z}^{(D)}\simeq 52.6$ mT in an elongated sample that was
also obtained experimentally.\cite{mchughetal09prb}

For simplicty we consider a long crystal of cylindrical shape of length $L$
and radius $R$ with the symmetry axis $z$ along the easy axis, magnetized
with $\sigma _{z}=$ $\sigma _{z}(z).$ The latter assumption makes the
problem tractable numerically. In this case the reduced magnetic field along
the symmetry axis has the form\cite{garchu08prb,garchu09prl,gar09prb}
\begin{equation}
D_{zz}(z)=\int_{-\infty }^{\infty }dz^{\prime }\frac{2\pi \nu R^{2}\sigma
_{z}(z^{\prime })}{\left[ \left( z^{\prime }-z\right) ^{2}+R^{2}\right]
^{3/2}}-k\sigma _{z}(z),  \label{DzzCylinh}
\end{equation}
where
\begin{equation}
k\equiv 8\pi \nu /3-\bar{D}_{zz}^{(\mathrm{sph})}=4\pi \nu -\bar{D}_{zz}^{(%
\mathrm{cyl})}>0,  \label{kDef}
\end{equation}
$k=14.6$ for Mn$_{12}$ Ac. For other shapes such as elongated rectangular,
one obtains qualitatively similar expressions.\cite{gar09prb} Now the total
field is given by
\begin{equation}
B_{\mathrm{tot},z}(z)=B_{z}+B_{z}^{(D)}(z)=B_{z}+\frac{Sg\mu _{B}}{v_{0}}%
D_{zz}(z).  \label{Btotz}
\end{equation}

One of the dynamical equations of the model is the relaxation equation for
the metastable population $n(t,z)$
\begin{equation}
\frac{\partial n(t,z)}{\partial t}=-\Gamma \left( B_{\mathrm{tot}%
,z}(z),T(z)\right) \left[ n(t,z)-n^{(\mathrm{eq})}(T)\right].   \label{ndot}
\end{equation}
In Eq. (\ref{ndot}) $\Gamma \left( \ldots \right) $ is the relaxation rate
taking into account both thermal actication over the barrier and resonance
spin tunneling that is calculated from the density matrix equation. As $B_{%
\mathrm{tot},z}$ depends on $n(z)$ everywhere in the sample via Eqs. (\ref
{DzzCylinh}) and (\ref{sigmazAvrOverd}), this is an integro-differential
equation. In the sequel we will set $n^{(\mathrm{eq})}\Rightarrow 0$ that is
a good approximation for strong enough bias.\cite{garchu97prb}

\begin{figure}[t]
\unitlength1cm
\begin{picture}(11,5)
\psfig{file=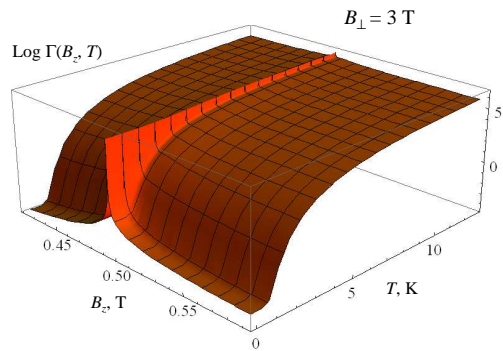,angle=-90,width=8cm}
\end{picture}
\caption{Relaxation rate $\Gamma (B_{z},T)$ in a generic model of a
molecular magnet.}
\label{Fig-Gamma}
\end{figure}

The second equation is the heat conduction equation that is convenient to
write in terms of the energy $\mathcal{E}$ of the system per unit cell as in
Ref. \onlinecite{garchu97prb} In the full-burning case $n^{(\mathrm{eq})}=0$
this equation has the form
\begin{equation}
\frac{\partial \mathcal{E}(t,z)}{\partial t}=\frac{\partial }{\partial z}%
\kappa \frac{\partial \mathcal{E}(t,z)}{\partial z}-n_{0}\Delta E\frac{%
\partial n(t,z)}{\partial t}.  \label{UnEqsDimensional}
\end{equation}
In Eq. (\ref{UnEqsDimensional}) $\kappa $ is the thermal diffusivity and $%
\Delta E$ is the energy released in the transition of a spin from the
metastable state to the ground state,
\begin{equation}
\Delta E=4hDS^{2},\qquad h\equiv \frac{g\mu _{B}B_{z}}{2DS}.
\label{DeltaEDef}
\end{equation}
The relation between the energy $\mathcal{E}$ and temperature is given by $%
\mathcal{E}(T)=\int_{0}^{T}C(T^{\prime })dT^{\prime },$ where $C(T)$ is the
experimentally measured heat capacity per unit cell.\cite{gometal98prb}

To solve the system of Eqs. (\ref{ndot}) and (\ref{UnEqsDimensional})
numerically, it is convenient to introduce reduced variables \cite
{garchu97prb}
\begin{equation}
\mathcal{\tilde{E}}\equiv \frac{\mathcal{E}}{n_{0}\Delta E},\qquad \tau
\equiv t\Gamma _{f},\qquad \mathbf{\tilde{r}}\equiv \frac{\mathbf{r}}{l_{d}},
\label{RedVars}
\end{equation}
where $n_{0}\leq 1$ is the initial population of the metastable state and $%
\Gamma _{f}$ is the relaxation rate at the flame temperature $T_{f}$ defined
by the energy balance $n_{0}\Delta E=\mathcal{E}(T_{f})$ and some fixed
value of $B_{\mathrm{tot},z}$ that we set to the resonance field $B_{k}.$
The characteristic distance $l_{d}=\sqrt{\kappa _{f}/\Gamma _{f}}$ defines
the width of the deflagration front in the case of normal (thermal)
deflagration and $\kappa _{f}$ is the thermal diffusivity at $T_{f}$ . In
terms of these variables, Eqs. (\ref{ndot}) and (\ref{UnEqsDimensional})
become
\begin{eqnarray}
\frac{\partial \mathcal{\tilde{E}}}{\partial \tau } &=&\frac{\partial }{%
\partial \tilde{z}}\tilde{\kappa}\frac{\partial \mathcal{\tilde{E}}}{%
\partial \tilde{z}}-\frac{\partial n}{\partial \tau }  \label{EtilEq} \\
\frac{\partial n}{\partial \tau } &=&-\tilde{\Gamma}\left( B_{\mathrm{tot}%
,z},T(\mathcal{\tilde{E}})\right) n,  \label{ntilEq}
\end{eqnarray}
where $\tilde{\Gamma}\equiv \Gamma /\Gamma _{f}$ is the reduced relaxation
rate and $\tilde{\kappa}\equiv \kappa /\kappa _{f}$. It remains to add the
expression for $B_{\mathrm{tot},z}$ in reduced variables, Eq. (\ref{Btotz})
with $D_{zz}(\tilde{z})$ given by Eq. (\ref{DzzCylinh}) with $z\Rightarrow $
$\tilde{z}$ and $R\Rightarrow \tilde{R}\equiv R/l_{d}$. The important
parameter $\tilde{R}$ is the ratio of the width of the front of tunneling
that is of order $R$ (see Refs. \onlinecite{garchu09prl,gar09prb}) to the
width of the standard deflagration front $l_{d}$. \cite
{suzetal05prl,garchu97prb}

Eqs. (\ref{EtilEq}) and (\ref{ntilEq}) are solved numerically by choosing a
finite-length sample and discretizing the problem in $\tilde{z}.$ This
yields a system of ordinary differential equations in time. We set $\tilde{%
\kappa}=1$ for simplicity. Before solving the equations, $\tilde{\Gamma}$
was calculated from the density matrix equation\cite{gar08-DME} for the
transverse field $B_{\bot }=3$ T and tabulated as a function of $B_{\mathrm{%
tot},z}$ and $\mathcal{E.}$ As $\tilde{\Gamma}$ increases by many orders of
magnitude near tunneling resonances, one has to use many different values of
$B_{\mathrm{tot},z}$ for interpolation here. In Fig. \ref{Fig-Gamma} one can
see that for such a strong transverse field the barrier is reduced to zero
at resonance where $\Gamma $ practically does not depend of temperature.
Thus near the resonance the cold deflagration should dominate, while off
resonance the regular deflagration should take place.

For the discussion it is convenient to consider the energy bias $%
W=\varepsilon _{-S}-\varepsilon _{m^{\prime }}$ between the two resonant
levels,
\begin{equation}
W=\left( S+m^{\prime }\right) g\mu _{B}\left( B_{z}+B_{z}^{(D)}-B_{k}\right)
\equiv W_{\mathrm{ext}}+W^{(D)}.  \label{WiDef}
\end{equation}
It is convenient to
use the reduced external bias
\begin{equation}
\widetilde{W}_{\mathrm{ext}}\equiv \frac{W_{\mathrm{ext}}}{2E_{D}}=\left( 1+%
\frac{m^{\prime }}{S}\right) \frac{v_{0}}{2Sg\mu _{B}}\left(
B_{z}-B_{k}\right) ,  \label{WextDef}
\end{equation}
where $E_{D}\equiv \left( Sg\mu _{B}\right) ^{2}/v_{0}$ is the dipolar
energy, $E_{D}/k_{B}=0.0671$ K for Mn$_{12}$ Ac. At the right end of the
dipolar window of Eq. (\ref{DipolarWindow}) one has $B_{z}=B_{k}+B^{(D)}.$
Thus with the help of Eq. (\ref{BviaD}) one obtains $\widetilde{W}_{\mathrm{%
ext}}=(1/2)\left( 1+m^{\prime }/S\right) D_{zz},$ i.e., $\widetilde{W}_{%
\mathrm{ext}}\approx D_{zz}$ for small bias, $m^{\prime }\approx S.$ We will
see that in the case of strong tunneling the speed of the quantum
deflagration front has a maximum at the right end of the dipolar window, $%
\widetilde{W}_{\mathrm{ext}}\approx \bar{D}_{zz}^{(\mathrm{cyl})}=10.53$ for
Mn$_{12}$ Ac.

\begin{figure}[t]
\unitlength1cm
\begin{picture}(11,5.5)
\psfig{file=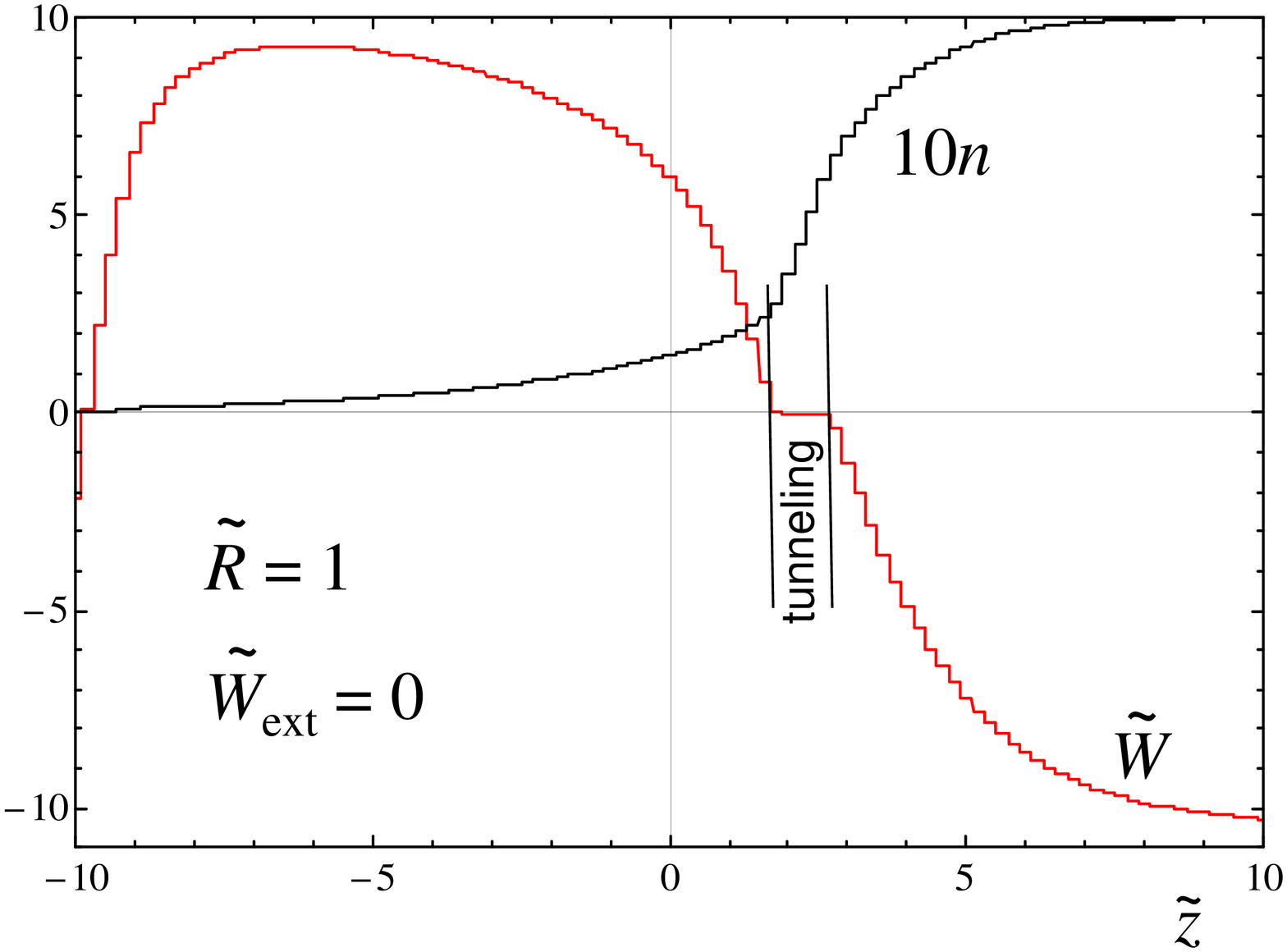,angle=-90,width=8cm}
\end{picture}
\begin{picture}(11,5.5)
\psfig{file=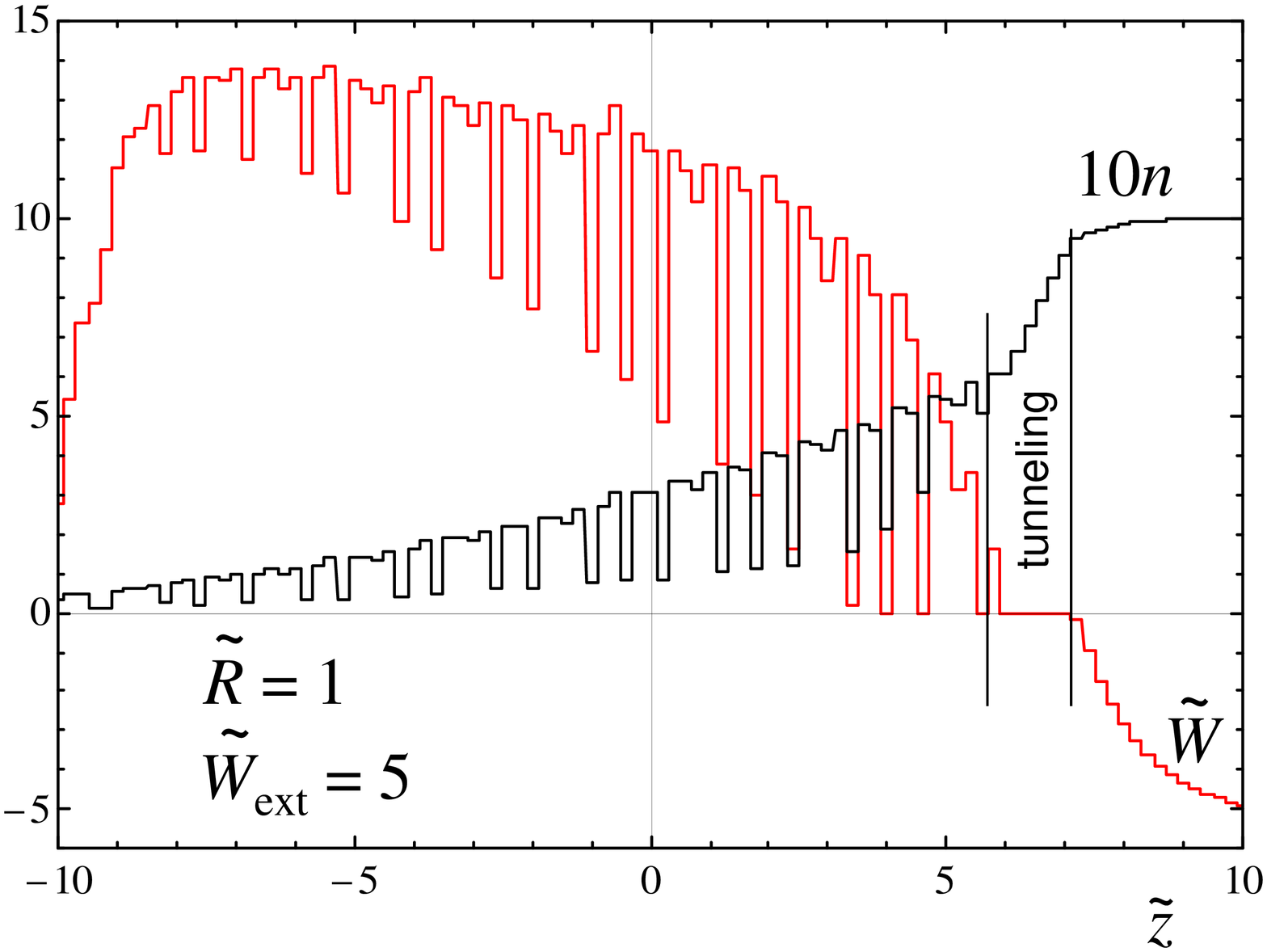,angle=-90,width=8cm}
\end{picture}
\caption{Profiles of the metastable population $n$ and the total bias $%
\widetilde{W}$ across the front for two values of the external bias $%
\widetilde{W}_{\mathrm{ext}}$: (a) $\widetilde{W}_{\mathrm{ext}}=0,$ laminar
regime; (b) $\widetilde{W}_{\mathrm{ext}}=5,$ non-laminar regime}
\label{Fig-Profiles}
\end{figure}

Cold deflagration can be ignited by the field sweep across the resonance. In
this case ignition occurs around the ,,magic'' value $\widetilde{W}_{%
\mathrm{ext}}\cong 5$ that corresponds to $B_{z}-B_{k}\cong 22$ mT.\cite
{garchu09prl,gar09prb} Outside the dipolar window fronts of tunneling do not
exist. On the other hand, standard deflagration can be initiated, at any
bias, by a quick temperature rise on one side of the sample. \cite
{garchu97prb} Applying this metnod of ignition here, we will see that within
the dipolar window the process is modified by spin tunneling and the speed
of the burning front can significantly increase for $\tilde{R}\gtrsim 1$,
especially at the right end of the window.

There are two regimes of propagation of non-thermal fronts of tunneling:
Laminar and non-laminar. Laminar regime with a smooth front takes place in
the left part of the dipolar window, $0\leq B_{z}-B_{k}\leq 10$ mT (or $%
0\leq \widetilde{W}_{\mathrm{ext}}\leq 1.3$), while the non-laminar regime
with frozen-in quasiperiodic spatial patterns of the magnetization behind
the front is realized in the right part of the dipolar window. In both
regimes burning is not complete and becomes less complete with increasing
the bias. In the laminar regime the residual magnetization and the front
speed were calculated analytically. \cite{gar09prb} The front speed
increases with the bias$.$ In the non-laminar regime, quasi-periodic
frozen-in patterns of magnetization deteriorate the resonance condition, and
the front speed decreases with the bias after the breakdown of the laminar
regime (see Fig. 5 of Ref. \onlinecite{gar09prb}). Thermal mechanism of
deflagration leads to complete burning of this residual metastable
population that smoothens the dipolar field profile in the sample and
improves the resonance condition inside the front core. This leads to the
increase of the front speed because of spin tunneling in the whole dipolar
window.

Results of numerical calculations for the spatial profiles of the metastable
population $n$ and the total bias $\widetilde{W}$ in the front for $\tilde{R}%
=1$ are shown in Fig. \ref{Fig-Profiles}. Both in the laminar and
non-laminar regimes, there is a region where $\widetilde{W}\cong 0$ and
resonant tunneling takes place, causing a greater slope of $n(z).$ Behind
the frond (on the left) metastable population $n$ burns to zero via thermal
mechanism.

\begin{figure}[t]
\unitlength1cm
\begin{picture}(11,5.5)
\psfig{file=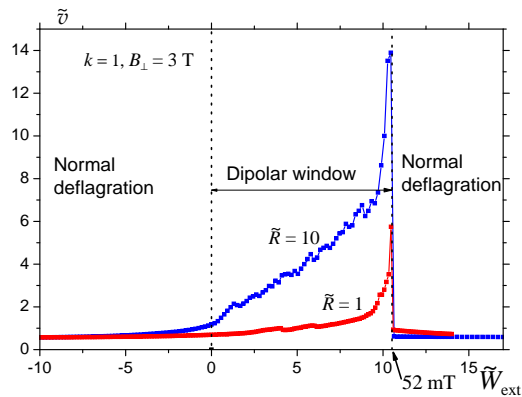,angle=-90,width=8cm}
\end{picture}
\caption{Reduced front speed $\tilde{v}$ vs bias field for different values
of the reduced transverse size $\widetilde{R}\equiv R/l_{d},$ $l_{d}$ being
the width of the thermal deflagration front. For such a strong applied
transverse field, the effect of tunneling is dramatic. }
\label{Fig-vTilde_vs_bias-generic}
\end{figure}

Numerical solutions for the reduced front speed $\tilde{v}=v/(l_{d}\Gamma
_{f})$ (Ref. \onlinecite{garchu07prb}) for the generic model with $B_{\bot }=3$ T
and $\tilde{R}=1$ and $10$ are shown in Fig. \ref{Fig-vTilde_vs_bias-generic}%
. Within the dipolar window the front speed can largely exceed the speed of
regular deflagration and depends on the transverse crystal size $R$
parametrized by $\tilde{R}.$ At $B_{\bot }=3$ T the maximal values of $%
\tilde{v}$ are attained at the right end of the dipolar window, followed by
a steep drop towards the standard-deflagration result outside the dipolar
window. For smaller transverse fields such as 2 T, the effect of spin
tunneling is weaker and $\tilde{v}$ \ reaches a maximum somewhere in the
middle of the dipolar window, depending on $\tilde{R}.$

Measurements of the speed of deflagration fronts \cite
{suzetal05prl,heretal05prl} were done in zero or small transverse field, so
that the influence of resonance spin tunneling on the front speed is not so
dramatic as in Fig. \ref{Fig-vTilde_vs_bias-generic}. It would be highly
interesting to perform deflagration experiments in strong enough transverse
field to see the big effect of tunneling on the front propagation. Changing
thermal contact of the crystal with the environment, one can boost or
suppress the thermal mechanism of magnetic burning thus isolating thermal
and quantum effects from each other.

Numerous useful discussions with E. M. Chudnovsky are greatfully
acknowledged.

This work has been supported by the NSF Grant No. DMR-0703639.

\bibliographystyle{apsrev}
\bibliography{chu-own,gar-own,gar-oldworks,gar-relaxation,gar-tunneling,gar-books,gar-spin}

\begin{thebibliography}{16}
\expandafter\ifx\csname natexlab\endcsname\relax\def\natexlab#1{#1}\fi
\expandafter\ifx\csname bibnamefont\endcsname\relax
  \def\bibnamefont#1{#1}\fi
\expandafter\ifx\csname bibfnamefont\endcsname\relax
  \def\bibfnamefont#1{#1}\fi
\expandafter\ifx\csname citenamefont\endcsname\relax
  \def\citenamefont#1{#1}\fi
\expandafter\ifx\csname url\endcsname\relax
  \def\url#1{\texttt{#1}}\fi
\expandafter\ifx\csname urlprefix\endcsname\relax\def\urlprefix{URL }\fi
\providecommand{\bibinfo}[2]{#2}
\providecommand{\eprint}[2][]{\url{#2}}

\bibitem[{\citenamefont{Friedman et~al.}(1996)\citenamefont{Friedman, Sarachik,
  Tejada, and Ziolo}}]{frisartejzio96prl}
\bibinfo{author}{\bibfnamefont{J.~R.} \bibnamefont{Friedman}},
  \bibinfo{author}{\bibfnamefont{M.~P.} \bibnamefont{Sarachik}},
  \bibinfo{author}{\bibfnamefont{J.}~\bibnamefont{Tejada}}, \bibnamefont{and}
  \bibinfo{author}{\bibfnamefont{R.}~\bibnamefont{Ziolo}},
  \bibinfo{journal}{Phys. Rev. Lett.} \textbf{\bibinfo{volume}{76}},
  \bibinfo{pages}{3830} (\bibinfo{year}{1996}).

\bibitem[{\citenamefont{Garanin and Chudnovsky}(1997)}]{garchu97prb}
\bibinfo{author}{\bibfnamefont{D.~A.} \bibnamefont{Garanin}} \bibnamefont{and}
  \bibinfo{author}{\bibfnamefont{E.~M.} \bibnamefont{Chudnovsky}},
  \bibinfo{journal}{Phys. Rev. B} \textbf{\bibinfo{volume}{56}},
  \bibinfo{pages}{11102} (\bibinfo{year}{1997}).

\bibitem[{\citenamefont{Chudnovsky and Garanin}(1997)}]{chugar97prl}
\bibinfo{author}{\bibfnamefont{E.~M.} \bibnamefont{Chudnovsky}}
  \bibnamefont{and} \bibinfo{author}{\bibfnamefont{D.~A.}
  \bibnamefont{Garanin}}, \bibinfo{journal}{Phys. Rev. Lett.}
  \textbf{\bibinfo{volume}{79}}, \bibinfo{pages}{4469} (\bibinfo{year}{1997}).

\bibitem[{\citenamefont{Garanin et~al.}(1998)\citenamefont{Garanin, Hidalgo,
  and Chudnovsky}}]{garmarchu98prb}
\bibinfo{author}{\bibfnamefont{D.~A.} \bibnamefont{Garanin}},
  \bibinfo{author}{\bibfnamefont{X.~M.} \bibnamefont{Hidalgo}},
  \bibnamefont{and} \bibinfo{author}{\bibfnamefont{E.~M.}
  \bibnamefont{Chudnovsky}}, \bibinfo{journal}{Phys. Rev. B}
  \textbf{\bibinfo{volume}{57}}, \bibinfo{pages}{13639} (\bibinfo{year}{1998}).

\bibitem[{\citenamefont{Garanin and Chudnovsky}(1999)}]{garchu99prb}
\bibinfo{author}{\bibfnamefont{D.~A.} \bibnamefont{Garanin}} \bibnamefont{and}
  \bibinfo{author}{\bibfnamefont{E.~M.} \bibnamefont{Chudnovsky}},
  \bibinfo{journal}{Phys. Rev. B} \textbf{\bibinfo{volume}{59}},
  \bibinfo{pages}{3671} (\bibinfo{year}{1999}).

\bibitem[{\citenamefont{Garanin}(2008)}]{gar08-DME}
\bibinfo{author}{\bibfnamefont{D.~A.} \bibnamefont{Garanin}},
  \bibinfo{journal}{arXiv:0805.0391}  (\bibinfo{year}{2008}).

\bibitem[{\citenamefont{Chudnovsky}(2004)}]{chu04prl}
\bibinfo{author}{\bibfnamefont{E.~M.} \bibnamefont{Chudnovsky}},
  \bibinfo{journal}{Phys. Rev. Lett.} \textbf{\bibinfo{volume}{92}},
  \bibinfo{pages}{120405} (\bibinfo{year}{2004}).

\bibitem[{\citenamefont{Chudnovsky et~al.}(2005)\citenamefont{Chudnovsky,
  Garanin, and Schilling}}]{chugarsch05prb}
\bibinfo{author}{\bibfnamefont{E.~M.} \bibnamefont{Chudnovsky}},
  \bibinfo{author}{\bibfnamefont{D.~A.} \bibnamefont{Garanin}},
  \bibnamefont{and}
  \bibinfo{author}{\bibfnamefont{R.}~\bibnamefont{Schilling}},
  \bibinfo{journal}{Phys. Rev. B} \textbf{\bibinfo{volume}{72}},
  \bibinfo{pages}{094426} (\bibinfo{year}{2005}).

\bibitem[{\citenamefont{{Y. Suzuki, M. P. Sarachik, E. M. Chudnovsky, S.
  McHugh, R. Gonzalez-Rubio, N. Avraham, Y. Myasoedov, E. Zeldov, H. Shtrikman,
  N. E. Chakov and G. Christou}}(2005)}]{suzetal05prl}
\bibinfo{author}{\bibnamefont{{Y. Suzuki, M. P. Sarachik, E. M. Chudnovsky, S.
  McHugh, R. Gonzalez-Rubio, N. Avraham, Y. Myasoedov, E. Zeldov, H. Shtrikman,
  N. E. Chakov and G. Christou}}}, \bibinfo{journal}{Phys. Rev. Lett.}
  \textbf{\bibinfo{volume}{95}}, \bibinfo{pages}{147201}
  (\bibinfo{year}{2005}).

\bibitem[{\citenamefont{{A. Hern\'andez-Minguez, J. M. Hern\'andez, F. Macia,
  A. Garcia-Santiago, J. Tejada, P. V. Santos}}(2005)}]{heretal05prl}
\bibinfo{author}{\bibnamefont{{A. Hern\'andez-Minguez, J. M. Hern\'andez, F.
  Macia, A. Garcia-Santiago, J. Tejada, P. V. Santos}}},
  \bibinfo{journal}{Phys. Rev. Lett.} \textbf{\bibinfo{volume}{95}},
  \bibinfo{pages}{217205} (\bibinfo{year}{2005}).

\bibitem[{\citenamefont{Garanin and Chudnovsky}(2007)}]{garchu07prb}
\bibinfo{author}{\bibfnamefont{D.~A.} \bibnamefont{Garanin}} \bibnamefont{and}
  \bibinfo{author}{\bibfnamefont{E.~M.} \bibnamefont{Chudnovsky}},
  \bibinfo{journal}{Phys. Rev. B} \textbf{\bibinfo{volume}{76}},
  \bibinfo{pages}{054410} (\bibinfo{year}{2007}).

\bibitem[{\citenamefont{Garanin and Chudnovsky}(2008)}]{garchu08prb}
\bibinfo{author}{\bibfnamefont{D.~A.} \bibnamefont{Garanin}} \bibnamefont{and}
  \bibinfo{author}{\bibfnamefont{E.~M.} \bibnamefont{Chudnovsky}},
  \bibinfo{journal}{Phys. Rev. B} \textbf{\bibinfo{volume}{78}},
  \bibinfo{pages}{174425} (\bibinfo{year}{2008}).

\bibitem[{\citenamefont{Garanin and Chudnovsky}(2009)}]{garchu09prl}
\bibinfo{author}{\bibfnamefont{D.~A.} \bibnamefont{Garanin}} \bibnamefont{and}
  \bibinfo{author}{\bibfnamefont{E.~M.} \bibnamefont{Chudnovsky}},
  \bibinfo{journal}{Phys. Rev. Lett.} \textbf{\bibinfo{volume}{102}},
  \bibinfo{pages}{097206} (\bibinfo{year}{2009}).

\bibitem[{\citenamefont{Garanin}(2009)}]{gar09prb}
\bibinfo{author}{\bibfnamefont{D.~A.} \bibnamefont{Garanin}},
  \bibinfo{journal}{Phys. Rev. B} \textbf{\bibinfo{volume}{80}},
  \bibinfo{eid}{014406} (\bibinfo{year}{2009}).

\bibitem[{\citenamefont{{S. McHugh, R. Jaafar, M. P. Sarachik, Y. Myasoedov, H.
  Shtrikman, E. Zeldov, R. Bagai, and G. Christou}}(2009)}]{mchughetal09prb}
\bibinfo{author}{\bibnamefont{{S. McHugh, R. Jaafar, M. P. Sarachik, Y.
  Myasoedov, H. Shtrikman, E. Zeldov, R. Bagai, and G. Christou}}},
  \bibinfo{journal}{Phys. Rev. B} \textbf{\bibinfo{volume}{79}},
  \bibinfo{pages}{052404} (\bibinfo{year}{2009}).

\bibitem[{\citenamefont{Gomes et~al.}(1998)\citenamefont{Gomes, Novak, Sessoli,
  Caneschi, and Gatteschi}}]{gometal98prb}
\bibinfo{author}{\bibfnamefont{A.~M.} \bibnamefont{Gomes}},
  \bibinfo{author}{\bibfnamefont{M.~A.} \bibnamefont{Novak}},
  \bibinfo{author}{\bibfnamefont{R.}~\bibnamefont{Sessoli}},
  \bibinfo{author}{\bibfnamefont{A.}~\bibnamefont{Caneschi}}, \bibnamefont{and}
  \bibinfo{author}{\bibfnamefont{D.}~\bibnamefont{Gatteschi}},
  \bibinfo{journal}{Phys. Rev. B} \textbf{\bibinfo{volume}{57}},
  \bibinfo{pages}{5021} (\bibinfo{year}{1998}).

\end{thebibliography}

\end{document}